\documentstyle[11pt,aaspp4]{article}
\begin{document}



\title{A comparison between Pa$\alpha$ and H$\alpha$ emission: 
the relation between mean \ion{H}{2} region reddening, local gas density and
metallicity\altaffilmark{1}}

\author{Alice C.~Quillen \& Mihoko Yukita}
\affil{Steward Observatory, The University of Arizona, Tucson, AZ 85721;
aquillen@as.arizona.edu} 

\altaffiltext{1}{Based on observations with the NASA/ESA 
{\it Hubble Space Telescope}
obtained at the Space Telescope Science Institute which is operated by the
Association of University for Research in Astronomy, Inc. (AURA), under
NASA contract NAS5-26555.}

\begin{abstract}
We measure reddenings to \ion{H}{2} regions in 
NGC~2903, NGC~1512, M51, NGC~4449 and NGC~6946 from Hubble Space Telescope
Pa$\alpha$ and H$\alpha$ images.
Extinctions range from $A_V \sim$ 5 -- 0 depending upon
the galaxy.  
For the galaxies with HST images in
both lines, NGC~2903, NGC~1512  and M51,
the Pa$\alpha$ and H$\alpha$ emission
are almost identical in morphology which implies
that little emission from bright \ion{H}{2} regions is hidden from
view by regions of comparatively high extinction. 
The scatter in the measured extinctions in each galaxy is only $\pm 0.5$ mag.

%
We compare the reddenings we measure in five galaxies 
using the Pa$\alpha$ to H$\alpha$ ratios
to those measured previously from the Balmer decrement
in the LMC and as a function of radius in M101 and M51.
We find that luminosity weighted mean extinctions of
these ensembles of \ion{H}{2} regions are
correlated with gas surface density and metallicity.  
The correlation is consistent with
the mean extinction depending on dust density 
where the dust to gas mass ratio scales with the metallicity.
This trend is expected if \ion{H}{2} regions tend to be located
near the mid-plane of a gas disk and emerge from
their parent molecular clouds soon after birth.
In environments with gas densities below a few hundred
M$_\odot$/pc$^2$,
star formation rates estimated from integrated line fluxes
and mean extinctions are likely to be fairly accurate.

\end{abstract}

\section {Introduction}

It has long been suspected that \ion{H}{2} regions tend to exhibit 
higher reddenings in higher metallicity environments.
For example, in  M101 and M51 (\cite{scowen3}; \cite{kennicutt96};
\cite{scowen_thesis}) a trend 
between \ion{H}{2} region extinction and galactic radius 
was observed and interpreted to be caused by the radial
metallicity gradient in these galaxies.
However, most studies have found a large scatter between 
the extinction and metallicity in individual \ion{H}{2} regions
(\cite{mccall}; \cite{zaritsky}; \cite{scowen}; \cite{kennicutt96};
\cite{bresolin}).  Even though the metallicity is fairly uniform
across the Large Magellanic Cloud (LMC),
high values of the Balmer decrement are measured 
near 30 Doradus (\cite{heydari}; \cite{caplan}) 
which is located in a region of high gas density 
(\cite{cohen}; \cite{luks}).
\cite{scowen} discovered a few \ion{H}{2} regions in M101 with anomalously 
high extinctions at large galactic radii, which they
attributed to higher local gas densities in nearby 
spiral arms.   
Despite the large scatter intrinsic to ensembles of \ion{H}{2} regions,
previous studies have found that the mean extinction of 
ensembles of \ion{H}{2} regions varied between different galaxy regions.
\cite{martin99} attributed the higher
mean extinctions in \ion{H}{2} regions in barred galaxies
compared to non-barred galaxies
to be caused by the higher gas densities associated with 
bar induced shocks. \cite{scowen} and \cite{scowen_thesis}
found that mean \ion{H}{2} region redennings were higher at 
large galactic radii than at smaller radii in both M101 and M51.

If there is a general trend for an ensemble of \ion{H}{2} regions to be
more highly reddened in higher gas density or metal rich environments,
then the star formation rate estimated from a sum of 
recombination line (such as H$\alpha$; e.g., \cite{kennicutt98})
fluxes would be underestimated in these environments.
One way to quantify this effect is to measure 
extinctions towards ensembles of \ion{H}{2} regions 
in a variety of different galactic environments.  
If mean \ion{H}{2} region extinctions 
depend upon gas density and metallicity 
then it might be possible to formulate an empirical correction for
the flux.
This would allow star formation rates to be measured 
from one recombination line in more heavily reddened regions.
If such a trend exists, star formation rates could be
more accurately estimated for galaxies at high redshift
and observed line ratios would also yield constraints
on the galaxy metallicity and gas surface density.

By observing in the the near-infrared it is possible
to probe into more heavily obscured regions because 
the near-infrared is less affected by dust extinction than optical 
wavelengths (for examples of previous work
on infrared recombination lines see \cite{calzetti96} and
\cite{beck}). 
The hydrogen Pa$\alpha$ recombination line at $1.875 \mu$m
is bright in the vicinity of hot newly formed stars.
It is difficult to observe in nearby galaxies
from the ground because of strong atmospheric absorption
from water molecules.  It is expected to be
about one eighth as bright as H$\alpha$ at temperatures
and densities typical of \ion{H}{2} regions, but since it is 
three times the wavelength of H$\alpha$ it should be less affected 
by extinction.  For an \ion{H}{2} region seen through a moderate foreground
dust screen with $A_V \sim 3$, Pa$\alpha$ should be as
bright as H$\alpha$.  Previous studies of Pa$\alpha$ imaging
in galaxies include \cite{boker} and \cite{marconi}.

In this paper we measure H$\alpha$ and Pa$\alpha$ fluxes for
\ion{H}{2} regions in the central regions of five galaxies.
We find that the scatter
in $A_V$ in each galaxy region is less than
the differences in the mean values.
Observations of distant galaxies are restricted towards
observing line emission over a large region in the galaxy.
Because of this we estimate a correction for the H$\alpha$
flux which is based upon summing over the extinction corrected
fluxes of individual \ion{H}{2} regions.
This procedure is one way of averaging over the scatter in extinctions 
which has been observed in previous studies of \ion{H}{2} regions.
We then investigate the dependence of this mean extinction on
local gas density and metallicity.


\section {Observations}

We searched the HST archive for galaxies which have high quality
narrow band images in Pa$\alpha$ and reveal a number
of \ion{H}{2} regions.  The archived images are listed
with their filters in Table 1.
NGC~2903 and NGC~1512 have narrow band images taken by HST in both Pa$\alpha$ 
and H$\alpha$, revealing a large number of \ion{H}{2} regions. 
M51 also has both Pa$\alpha$ and H$\alpha$ 
images observed by HST showing numerous 
\ion{H}{2} regions; however, these images overlap only in a small region
about $20''$ to the north and $20''$ to the east of the nucleus.  
NGC~4449 and NGC~6946 
have high quality Pa$\alpha$ images containing numerous \ion{H}{2} regions but
lack HST narrow band imaging in H$\alpha$.
For comparison we used H$\alpha$ fluxes for NGC~4449 measured
by \cite{scowen_thesis} and 
a calibrated ground-based H$\alpha$ image for NGC~6946 provided
by Rob Kennicutt (private communication).
NGC~2903, NGC~4449 and NGC~6946 were observed as part 
of the nearby galaxies snap shot survey (\cite{boker}).

The NICMOS images were reduced with the {\sl nicred} data reduction
software (\cite{mcl}) using on orbit darks and flats.  
The narrow band images required a pedestal correction
which we assumed was an additional constant in each
quadrant of the chip 
(for more discussion on this see \cite{boker}).
WFPC2 images were combined using the IRAF/STSDAS routine
{\sl combine} which removed most of the cosmic ray hits.
We derived H$\alpha$ and Pa$\alpha$ line emission images by
subtracting scaled continuum images from the narrow band
images containing the lines.

WFPC2 images were calibrated using zeropoints listed in
the HST Data Handbook.  Specifically we used 
$3.062\times 10^{-15}$ and
$2.941\times 10^{-15}$ 
erg~cm$^{-2}$~s$^{-1}$/(DN~s$^{-1}$)
for the continuum subtracted F656N and F658N narrow band images,
respectively in the PC camera of WFPC2.
The NICMOS continuum subtracted narrow band images in the F187N
filter were calibrated using
$6.69\times 10^{-16}$ and 
$8.27\times 10^{-16}$ erg~cm$^{-2}$~s$^{-1}$/(DN~s$^{-1}$)
for Camera 2 and Camera 3 respectively
for the Pa$\alpha$ line. 
The NICMOS broad-band images in the F160W filter were calibrated using
$2.190\times 10^{-6}$ and 
$2.776\times 10^{-6}$ Jy/(DN s$^{-1}$) for Camera 2 and Camera 3 respectively.
The NICMOS flux calibration is 10.7\% lower than that used
by \cite{boker} and is based on
measurements of the standard stars P330-E and P172-D during
the Servicing Mission Observatory Verification program
and subsequent observations (M.~Rieke 1999, private communication).
The filter widths listed in the NICMOS and WFPC2 instrument handbooks
were used to derive the calibration corrections.
The location of the lines was near the region of maximum
transmission in the filters
in all cases so no correction for filter transmission 
is required. 


To measure the flux of emission regions we used aperture photometry.  We 
identify individual regions from peaks in the Pa$\alpha$ line emission images.
Aperture sizes were chosen to match the resolution of the H$\alpha$ images 
or that of existing H$\alpha$ photometry 
(for NGC 4449 to match that of \cite{scowen_thesis}).
For NGC~2903, NGC~1512, and M~51, because we had
HST imaging in both lines we were able to use very small apertures.
In all cases identical aperture positions and sizes were used
on both line emission images.
Fluxes and identified regions are listed in Tables 2-6.
Positions are given with respect to the nucleus as defined
by the centroid at the brightest spot in the NICMOS/F160W images.
The positions of nuclei are listed in Table 2-6.
Astrometry was done with the IRAF/STSDAS routine {\sl xy2rd} 
from the F160W images
and is that determined by the HST pointing.


We use the H$\alpha$/Pa$\alpha$ line ratios to estimate the extinction
to each region.   We assume intrinsic flux ratios of
$F_{H\alpha}/F_{H\beta} = 2.86$,
$F_{H\alpha}/F_{Pa\alpha} = 8.46$ for Case B recombination
at a temperature of $10^4$K and a density $100$ cm$^{-3}$
(line ratios are given in \cite{osterbrock})
and a Galactic extinction law (\cite{mathis}).                          
For NGC~2903 and M51 the [NII]6584$\AA$ line was not contained
in the F656N filter,  however it is contained in the F658N
image of NGC~1512.  We correct for the presence of this line
in NGC~1512 and NGC~6946
assuming a H$\alpha$/(H$\alpha$ + [NII]) line ratio of 0.75
(following the procedure of \cite{kennicutt83}).
We corrected for Galactic redenning using values from \cite{schlegel}.

Fluxes and extinctions derived from Pa$\alpha$/H$\alpha$ flux ratios are 
listed in Tables 2-6.  The variations in the intrinsic line ratios 
caused by temperature differences 
are unlikely to significantly affect the measured extinctions;
(lowering the electron temperature by 5000K results in a reduction of 
of 15\% in the H$\alpha$/Pa$\alpha$ line ratio).
The errors in the measured line ratios are dominated
by errors in the sky value caused by color variations
which affect our continuum subtraction or the uneven sky 
in the Pa$\alpha$ images caused by the pedestal affect.  
The photometric uncertainties are roughly estimated 
in the Table captions.  There is an additional error
in the absolute calibration of $\sim \pm 20\%$ which corresponds
to an error of $\sim \pm 0.2$ in the derived $A_V$.
Relative photometric error between individual \ion{H}{2} 
regions in a galaxy
are $\sim \pm 0.1$ mag in $A_V$ for the faintest regions.
The scatter, $\sim \pm 0.5$, in $A_V$ in each galaxy
is therefore likely
to be intrinsic and not caused by photometric uncertainty.


When possible, we have compared our measured extinction values with those
reported in previous studies.
Two of the \ion{H}{2} regions in NGC 6946 were also studied
by  \cite{hyman} with radio continuum and H$\alpha$ observations.
For the 3rd and 4th \ion{H}{2} region listed in Table 5 we find
$A_V \sim 3.2$, while \cite{hyman} report $A_V \sim 2.5$ and $3.4$,
in rough agreement with our estimates.

Although \cite{scowen_thesis}
measured a high extinction, $A_V \sim 2$, for many regions in NGC~4449 from his
H$\alpha$ and H$\beta$ narrow band imaging, the Pa$\alpha$/H$\alpha$ 
line ratios that we compute from our Pa$\alpha$ photometry and his H$\alpha$ 
fluxes suggest that there is little internal extinction in this galaxy.
However, our line ratios are consistent with the spectroscopic measurements of
\cite{kobulnicky} who find H$\alpha$/H$\beta$ $\sim 2.86$ which 
means there is little extinction.   We suspect that there may have been an
error in the calibration of the H$\beta$ image by \cite{scowen_thesis}.
H$\alpha$ fluxes listed by \cite{fuentes} are consistent with those measured
by \cite{scowen_thesis} so 
the H$\alpha$ fluxes measured by \cite{scowen_thesis} are sufficiently
well-calibrated.
It is a concern that extinctions measured
from the Balmer decrement might be biased by 
H$\beta$ absorption present in the stellar component 
(e.g., \cite{dufour}).
However, after correcting for H$\beta$ absorption from the stellar
component, \cite{scowen_thesis} found   
no strong changes in his correlation plots.   

\subsection {Comparing H$\alpha$ and Pa$\alpha$ Morphology}

In Figures 1-3 we show the HST images for NGC~1512, NGC~2903, M~51
in both Pa$\alpha$ and H$\alpha$ lines.
We found little correspondence between star clusters observed
in the broad band images and the line emission maps.
This makes it difficult to constrain the ages of the cluster
based on Pa$\alpha$ equivalent widths.
A number of star clusters were observed 
which were outside the region of line emission emission and so could
correspond to regions where radiation pressure, 
ionization fronts, winds and supernovae have cleared the clusters of gas.
In NGC~1512 and NGC~4449 the luminosities in H$\alpha$ of the faintest regions 
are $\sim 10^{37}$ ergs/s and so similar in luminosity to the
Orion nebula (\cite{kennicutt84}) which contains one O star. 
Since \ion{H}{2} regions cannot be much fainter than this, we suspect
that there are no deeply embedded \ion{H}{2} regions in these two galaxies.
The faintest regions detected in NGC 2903 and M51 are an order
of magnitude brighter and so could contain a few dozen 
O stars.  

In NGC~1512, NGC~2903 and M51, there are few qualitative differences between
the respective line emission images.  In other words,  
almost all \ion{H}{2} regions above the detection limit in Pa$\alpha$ 
were seen in both images.
One region stands out as an exception to the above statement.
This region in NGC 2903 is
$0\farcs 71$ east and $4\farcs 17$ north of the nucleus (see Figure 1) and
is significantly brighter in Pa$\alpha$ 
than in H$\alpha$.  From the line ratio we estimate 
$A_V \sim 4$, however we did not see a peak in the H$\alpha$
line image; instead the line emission is diffuse and may not
be emitted from the same spatial region.
Previous studies have measured large extinctions
in the nucleus of this galaxy ($A_V \sim 15$; \cite{lebofsky}) as well
as high gas densities (\cite{jackson}) so this region may be an 
example of an obscured region.

Total Pa$\alpha$ and H$\alpha$ fluxes across our NGC~2903 
images are $4.5\times 10^{-13}$
and $5.3\times 10^{-13}$ ergs cm$^{-2}$s$^{-1}$ respectively.  
We can compare these values
to the Br$\alpha$ and Br$\gamma$ fluxes measured by \cite{beck} in an $8''$
diameter aperture.  We would predict 
Br$\alpha$ and Br$\gamma$ fluxes of $1.53\times 10^{-13}$, and 
$5.30\times 10^{-14}$
ergs cm$^{-2}$s$^{-1}$ respectively, using an $A_V$ of 3.06 consistent 
with the Pa$\alpha$ and H$\alpha$ line ratio across this aperture.
Our estimated Br$\alpha$ and Br$\gamma$ fluxes are consistent 
with those measured by \cite{beck}, 
$1.4 \pm 0.18 \times 10^{-13}$   and
$3.5 \pm 1.2 \times 10^{-14}$ respectively. 
The extinction we estimate $A_V \sim 3-4$ is therefore consistent with 
line fluxes and ratios
measured by \cite{beck},  but substantially lower than that, $A_V \sim 15$,  
measured from broad and narrow band imaging near the 10$\mu$m 
silicate absorption feature (\cite{lebofsky}).
The region identified above is not bright enough (it is only 1/20th the total) 
in Pa$\alpha$ to dominate the total
Pa$\alpha$ emission in the $8''$ nuclear region.  
Since the line ratios agree with $A_V \sim 3-4$,
it is also unlikely that this region dominates the Br$\alpha$ emission.
The total 10$\mu$m flux roughly corresponds to that
expected if a moderate fraction (e.g. 1/2) of
the total bolometric luminosity in the UV (needed 
to account for the line emission) 
is re-radiated by dust (\cite{beck}).   
The $10\mu$m emission is most likely to be emitted from the extended
region we see in the line emission maps and not from a single deeply
embedded region.  The most likely explanation for the discrepancy
between the extinctions we measure and that estimated at 10$\mu$m 
is that the $10\mu$m spectrum contains structure from 
by PAH emission which caused \cite{lebofsky} to overestimate the depth
of the silicate absorption feature (see for example \cite{siebenmorgen}).
The extinction we estimate near the nucleus of NGC~6946 
($A_V \sim  4.6$) is also significantly lower than that estimated
by \cite{lebofsky} ($A_V \sim 50$).

\section {A comparison between gas surface density, metallicity and
extinction}

For a distant galaxy we would like to 
to derive star formation rates based on the 
emission line flux.  We therefore desire a way to 
estimate the total emission in this line that would
be emitted in the absence of extinction.
In the comparison between H$\alpha$ and Pa$\alpha$
images we found no evidence for a population bright HII regions
that are extremely red compared to the average.  The scatter
in $A_V$ in each galaxy region (approximately $\pm 0.5$) is less than
the differences in the mean values which range from a mean 
$A_V \sim 3.3$ to $0$.   Because the mean values span
a large range, we are prompted to look for correlations
between this value and environmental properties.

To compare extinctions in different galactic environments we must first 
compile information from the literature 
on the metallicities and gas densities in the various 
regions of these galaxies.  These are listed in Table 7 with references
to the literature from which they were taken.
The gas densities are primarily taken from single
dish CO measurements with similar aperture sizes.
When necessary we have corrected published values of CO so that we used
the same conversion factor of
N(H$_2$)~$= 2.8\times 10^{20} {\rm cm}^{-2} I_{CO}$~(K~km~s$^{-1}$)
(Scoville et al.~1987).  
The metallicities were primarily taken from Zaritsky et al.~(1994).
We estimate that the uncertainties in the gas density 
(approximately a factor of 2) and
metallicity measurements (approximately $0.3$ dex) 
are smaller than the variations between
the measurements in different galaxies and galaxy regions.

We did not find a measured gas mass in NGC~1512 so we estimate
the gas surface density based on the color map made from the 
broad band V ($0.5\mu$m ) and H band ($1.6\mu$m) images.
We observed a change in color of $\Delta$(V-H)~$\sim 0.4$
between the circumnuclear ring and the region interior and exterior
to it.   We assume that the dust is a foreground screen 
for half of the light (for the bulge stars that are located behind
the gas).  This is equivalent to assuming
that the scale length of the gas is smaller than that
of the stars.
This change in color then corresponds to $A_V \sim 1.3$ and  
a gas surface density in hydrogen of 
$\Sigma_H \sim 19  M_\odot~{\rm pc}^{-2}$
(based on the $A_V$ to N(H) conversion and 
extinction law of \cite{mathis}).

We did not find any optical spectroscopy for NGC 1512 
in the literature.
However, we can use the correlation between absolute magnitude
and central metallicity found by \cite{garnett97}
to estimate the metallicity in its circumnuclear ring.
This galaxy has an absolute magnitude of $M_B \sim -18.9$ (assuming
a distance of 9.5 Mpc from \cite{tully} with a Hubble constant
of 75 km~s$^{-1}$ Mpc$^{-1}$).
The galaxy has an inclination corrected circular velocity
of $v_c \approx 173$ km/s.   From Figure 4 of \cite{garnett97}
we estimate that NGC~1512 should have a central metallicity
that is approximately solar.

To increase the size of our comparison sample
we also compile the properties of \ion{H}{2} regions
as a function of radius in M101 and M51.
\ion{H}{2} region extinctions and luminosities in M101 were taken
from \cite{scowen} and are restricted to 
the 200 brightest \ion{H}{2} regions.
To describe the azimuthal gas density in M101 we summed the 
azimuthal molecular and
atomic hydrogen gas mass densities reported by \cite{kenney}.
The metallicity gradient was taken from \cite{kennicutt96}.
\ion{H}{2} region extinctions and luminosities in M51 were taken
from \cite{scowen_thesis} and are restricted to be from
the brightest \ion{H}{2} regions defined as having an observed 
flux in H$\alpha$ greater than $10^{-14}$~erg~cm$^{-2}$~s$^{-1}$.
Molecular and atomic hydrogen gas densities 
in M51 were taken from  \cite{rydbeck} and \cite{scoville83}.
The oxygen abundance gradient in M51 were taken from \cite{zaritsky}.


\subsection{Weighting}

We approach the problem of correcting
for extinction by considering a sum over individual \ion{H}{2} regions
and by deriving a correction that can be multiplied
by the total observed flux in the line to obtain the true one.
This procedure is one way of averaging over a large scatter in extinctions
and allows us to look for trends between the mean extinction and 
environmental properties.
The correction we can describe as a luminosity weighted mean extinction.
To derive this correction we sum over the corrected fluxes
of each region, so it involves a luminosity weighted sum.   
This makes the estimate more robust than
using the mean extinction computed from the distribution 
of extinctions measured towards each region.  
The mean computed from the distribution
would be more dependent on the faintest
HII regions and so noisier and less robust than the luminosity
weighted mean extinctions computed as in equation (3) below.

We denote the flux in H$\alpha$ from a region $i$ by $F_{H\alpha,i}$.
We denote the sum of observed fluxes over a larger area 
as $F_{H\alpha} = \sum_i F_{H\alpha,i}$.
We denote the true or corrected emission as 
\begin{equation}
F_{H\alpha,corr} = \sum_i F_{H\alpha,i,corr}.
\end{equation}
We aim to estimate a factor $a_{H\alpha}$ such
that we can multiply the observed total flux in
a large region to obtain the true or corrected emission.
\begin{equation}
F_{H\alpha,corr} = a_{H\alpha}  F_{H\alpha}.
\end{equation}
We correct the fluxes from each region using the extinctions
derived from the line ratios and then sum the regions 
to estimate $F_{H\alpha,corr}$.  At the same
time we sum the uncorrected fluxes to get $F_{H\alpha}$.
From the ratio of $F_{H\alpha}$ and $F_{H\alpha,corr}$ 
we estimate the correction
factor $a_{H\alpha}$ that we need to obtain the true
or corrected emission from the observed total.
This is converted to a mean extinction at the wavelength
of H$\alpha$ by 
\begin{equation}
\overline{A(H_\alpha)} \equiv -2.5 \log{a_{H\alpha}}.
\end{equation}
A similar mean extinction can also be calculated for Pa$\alpha$.
This can be compared to the extinction estimated from the ratio of 
the total Pa$\alpha$ and H$\alpha$ fluxes which we 
refer to as $A_{total}(H\alpha)$.

We calculate the luminosity weighted mean extinction (defined above) by 
summing over the Pa$\alpha$ 
and H$\alpha$ fluxes listed in Tables 2-6
for the nuclear regions of NGC~1512, NGC~2903, NGC~4449,
M51 and NGC 6946.   These mean extinctions are listed
in Table 7.  We also perform these sums using
H$\alpha$ and H$\beta$ fluxes as a function of radius
in M101 and M51 based on the Balmer decrement.
The radial width over which we summed the properties 
of \ion{H}{2} regions was 1~kpc for M51 and 1.5 kpc for M101.
For comparison to an additional low metallicity system
we also derive a similar estimate for the LMC 
based on the Balmer decrement.
H$\alpha$ and H$\beta$ fluxes for LMC \ion{H}{2} regions
were taken from \cite{caplan}
and we correct for Galactic extinction using a value
of $A_B = 0.324$ (\cite{schlegel}).

In Table 7 we can compare $\overline{A(H\alpha})$ with $A_{total}(H\alpha)$.  
$\overline{A(H\alpha})$ allows us to correct the total
observed H$\alpha$ flux for redenning based on the reddenings 
to each \ion{H}{2} region.  
We see a tendency for $\overline{A(H\alpha})$
to be somewhat higher than that estimated from the line ratio 
of the integrated fluxes ($A_{total}(H\alpha)$).
The differences are small because the dispersion in the extinctions
between individual \ion{H}{2} regions in each galaxy is low.
A star formation rate estimated from the total H$\alpha$ line
flux and corrected with an extinction based on 
the H$\alpha$/Pa$\alpha$ line ratio is likely
to be quite accurate.

We plot our mean extinctions (defined in Eqn 3) versus gas density
in Figure 4a.   As expected there is a correlation between
mean extinction and local gas density.  However some of the
correlation may be due to the dependence of the CO to H$_2$ 
conversion factor on metallicity.
To take this dependence into account 
we correct the H$_2$ by a factor that depends on the metallicity
and display the corrected gas density in Figure 4b.
Our correction assumes an increase in the conversion factor by 4.6 when
the oxygen abundance drops by a factor of 10 (\cite{wilson95}).
Extending this relation towards metallicities larger
than solar is reasonable since 
galaxies with metallicities greater than solar are 
expected to have lower conversion factors (e.g., \cite{nakai95}).
The correlation between gas density and mean extinction is not significantly
changed.

It is difficult to separate between the role
of gas density and that of metallicity 
since the densest regions also correspond to the most metal rich.
However as we see in Figure 4c, metallicity alone
is not sufficient to account for the large variation in 
our mean extinctions.

The dust density is the quantity most likely to 
affect the reddening so we desire a comparison between
our extinction values and the dust surface density.  
We can estimate the dust surface density from the gas density
using a metallicity dependent gas to dust mass ratio. 
The gas to dust ratio may correlate better with
the carbon abundance than with the oxygen abundance
(\cite{bohlin}; \cite{martin89}).
To convert between oxygen abundance and carbon abundance
we use the relation measured by \cite{garnett95} 
[log(C/O)~$\approx 0.54$~log(O/H)~$+ 1.6$].
We assume that at 
solar metallicity the gas to dust mass ratio is 120.
To estimate the dust surface density we multiply
the solar dust to gas ratio by the gas density and the
ratio between the carbon abundance
predicted from the oxygen abundance and that at solar.
The relation between mean extinction and estimated dust density
is shown in Figure 4d.
The dashed line plotted in Figure 4d corresponds to that predicted
for a uniform medium with the \ion{H}{2} regions located
in the middle of the dust distribution.
We find that the correlations we have seen between mean extinction, 
gas density and 
metallicity are consistent with an approximately linear relation
between the log of the dust density and mean extinction.

For the high dust densities, 
the mean extinctions expected are somewhat higher than 
those observed.
In Figure 4d we plotted the extinction predicted if the \ion{H}{2}
regions were in the mid plane of a uniform disk.
These deviations would be predicted from a model
which takes into account the patchiness of the ISM
(e.g., \cite{calzetti96}).   To illustrate
this we have also plotted in Figure 4d a patchy dust model
for a two component dust model with an extinction ratio of four and
an area filling factor of 0.5.  However such a model
would result in a larger scatter (greater than $\pm 0.75$)
in the extinctions towards individual regions than we measured.
A more realistic patchy dust model may match the extinction and
mean gas density.  There could be nearby dense molecular clouds lacking
\ion{H}{2} regions which would have contributed to the mean gas
density in the region but not to the \ion{H}{2} region extinctions.

\section {Summary and Discussion}

In this paper we have presented a comparative study between
Pa$\alpha$ and H$\alpha$ in the central regions of five galaxies.
For three galaxies we have HST images in both lines and the
morphologies of the respective line emission maps are remarkably
similar.   The dispersion in extinctions is about 
a magnitude for the \ion{H}{2} regions in each galaxy and is smaller
than the differences between the mean extinctions.
We find no evidence for a population of bright
heavily embedded \ion{H}{2} regions. 

Only one bright region, in NGC 2903, was detected
clearly in the Pa$\alpha$ image and was faint or not
seen in the H$\alpha$ image.  However this region does
not dominate the Pa$\alpha$ flux of the nuclear star forming
region and is also unlikely to dominate the Br$\alpha$ emission.
This suggests that an estimate of the star formation rate
based on a mean extinction and either the H$\alpha$ or Pa$\alpha$
flux is likely to be fairly accurate.
We suspect that some previous studies of the silicate
absorption feature at 10$\mu$m may have overestimated
the extinction towards the nuclei of NGC 2903 and NGC 6946 because
of contamination by PAH features.

We have compared mean extinctions as a function of gas
density and metallicity and find correlations between
these quantities.  
The correlations between mean extinction, gas density and
metallicity are consistent with an approximately linear relation
between the log of the dust density and mean extinction.
This trend is expected if the \ion{H}{2} regions tend to be located
near the mid-plane of a gas disk and emerge from  
their parent molecular clouds soon after birth.
Radiation pressure and ionization fronts are expected to
clear a young star forming region of dense gas 
on a timescale of $10^5$ years
(\cite{ferland}) whereas \ion{H}{2} regions should exist for a few times
$10^{6}$ years.   It would be unlikely
to detect an \ion{H}{2} region in a region with extinction
significantly higher than the mean predicted from the local
gas density.  This may provide a partial explanation
for the low dispersion in extinctions that we measured in
each galaxy.

We expect that forthcoming high resolution CO and HI surveys will allow
a more careful comparison between the statistical 
properties of \ion{H}{2} regions
and the gas distribution.  It may be possible to study correlations 
between the patchy distribution of the ISM, the tendency for
star forming regions to evacuate themselves of dense gas,
and large scale gaseous structures cause by bars and spiral arms.
The relationships presented here may allow more
accurate estimates of star formation rates
in galaxy regions which contain gas densities 
up to a few hundred $M_\odot$ pc$^{-2}$.
Above this level it is likely that a significant fraction
of the line emission is absorbed even in the near-infrared and 
mid or far infrared measurements would be required (e.g., \cite{kennicutt98}).
Further work on the trends discussed here may make it possible
to more accurately measure star formation rates
and constrain gas densities in distant galaxies.


\acknowledgments

Support for this work was provided by NASA through grant number
GO-07869.01-96A
from the Space Telescope Institute, which is operated by the Association
of Universities for Research in Astronomy, Incorporated, under NASA
contract NAS5-26555.
We also acknowledge support from NASA project NAG-53359 and
NAG-53042 and from JPL Contract No.~961633.
This work could not have been done without the help of Don Garnett.
We also thank Rob Kennicutt, Marcia Rieke, George Rieke, Torsten B\"oker,
Chad Engelbracht and Karl Gordon for helpful discussions.
We thank Gil Rivlis for helping us scan and convert tables.
We thank Rob Kennicutt for providing us with the H$\alpha$
image of NGC~6946.
This research has made use of the NASA/IPAC Extragalactic Database (NED) 
which is operated by the Jet Propulsion
Laboratory, California Institute of Technology, under contract with 
the National Aeronautics and Space Administration. 
We thank the referee for comments which have improved this paper.

\clearpage


\begin{figure*}
\vspace{16.0cm}
\includegraphics{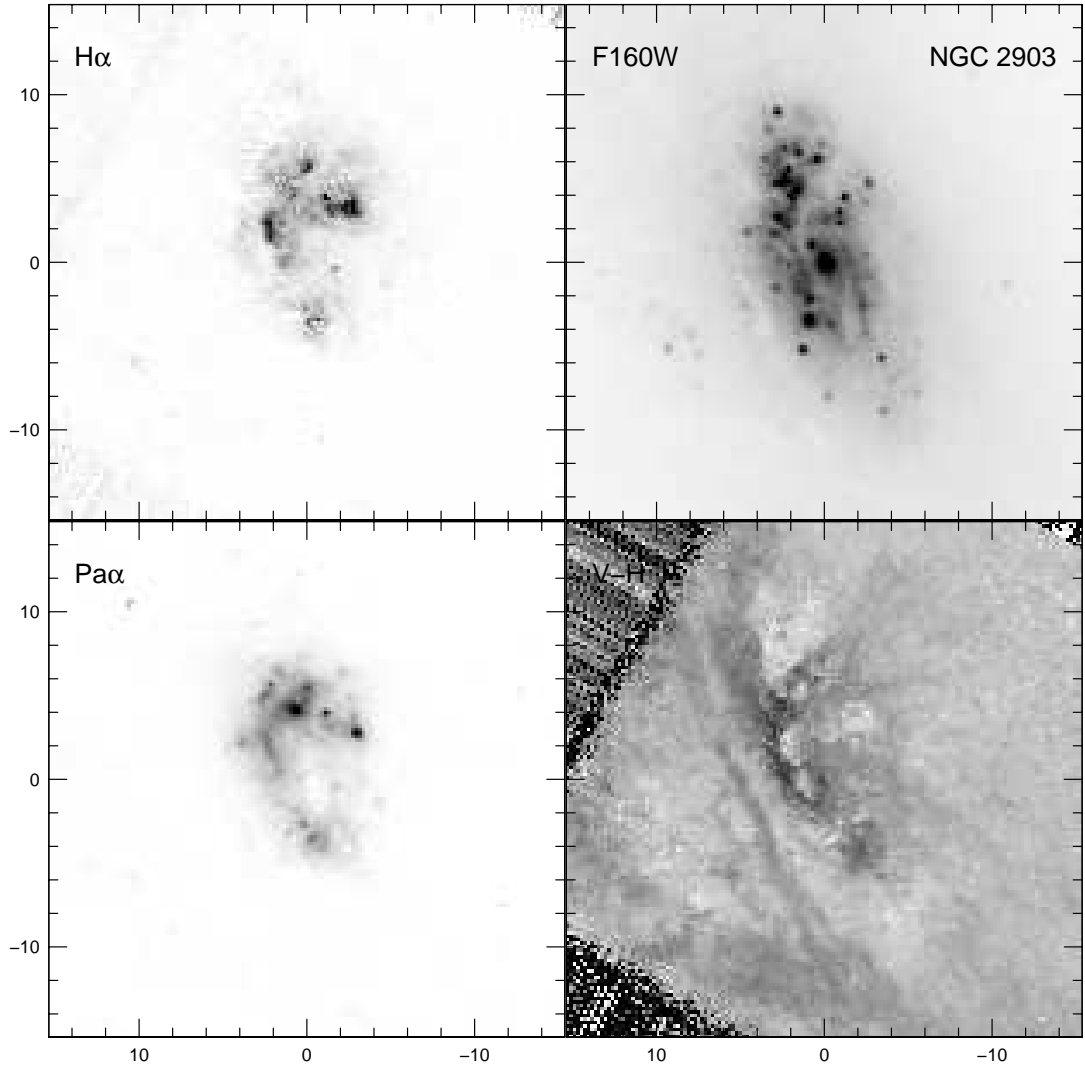}
\caption[junk]{
HST images of the circumnuclear ring of NGC 2903. The H$\alpha$
image was constructed WFPC2 images.  The Pa$\alpha$
was observed with NICMOS Camera 3 and constructed from F187N and F160W
images.  The V-H color map was constructed from the F555W and F160W
images.  Colors from V-H $= 2.0$ (white) to 3.5 (black) are shown.
Axes are given in arcsec from the nucleus.
There is a bright source at $0\farcs 71$ east, and $4\farcs 17$ north 
from the nucleus evident in
the Pa$\alpha$ image but not the H$\alpha$ image.
}
\end{figure*}

\begin{figure*}
\vspace{16.0cm}
\includegraphics{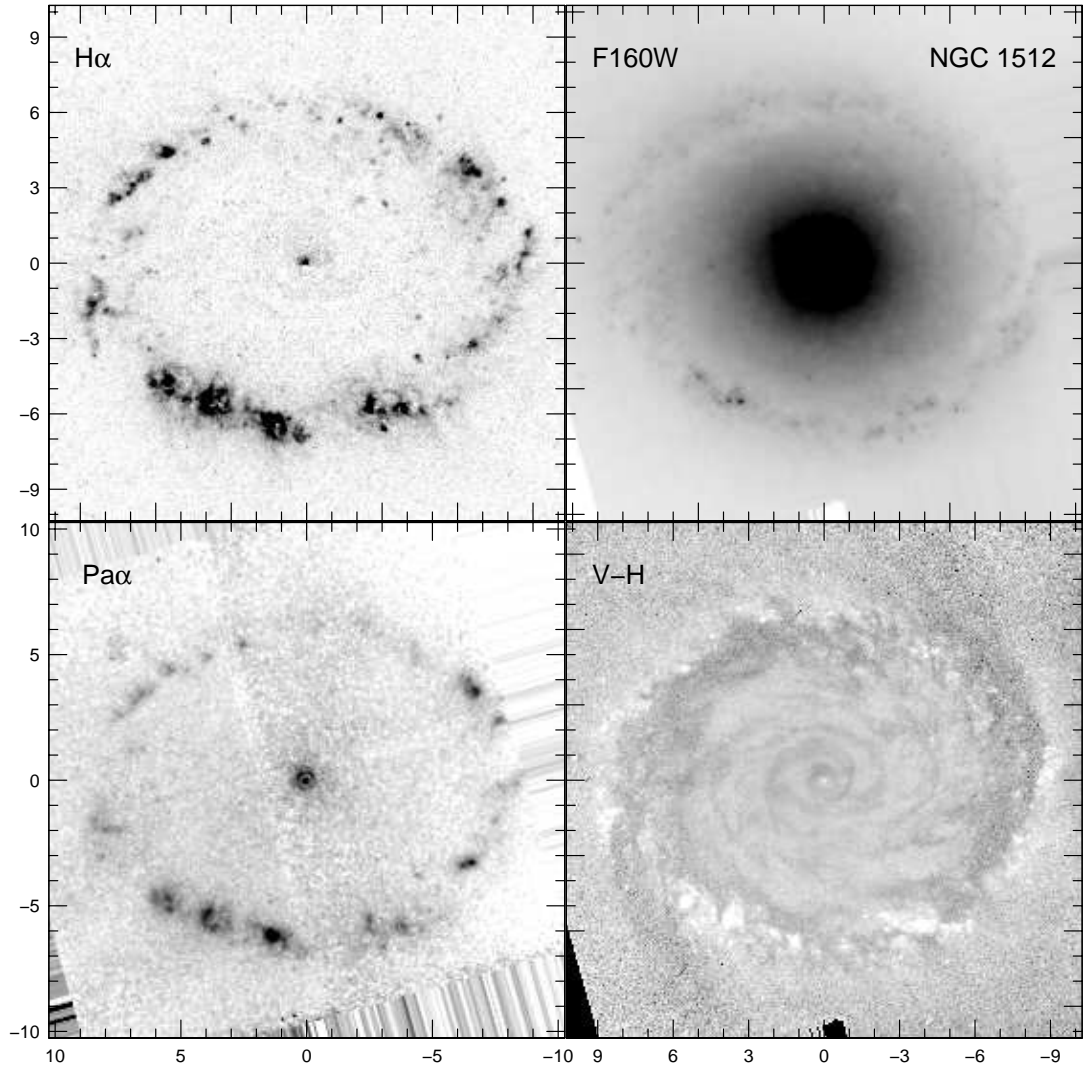}
\caption[junk]{
HST images of the circumnuclear ring of NGC 1512. The H$\alpha$
image was constructed WFPC2 images.  The Pa$\alpha$
was observed with NICMOS Camera 2 and constructed from F187N and F160W
images.  The V-H color map was constructed from the F547M and F160W
images.  Colors from V-H $= 2.5$ (white) to 3.3 (black) are shown.
Axes are given in arcsec from the nucleus.
We found no evidence for bright embedded \ion{H}{2} regions.
}
\end{figure*}

\begin{figure*}
\vspace{16.0cm}
\includegraphics{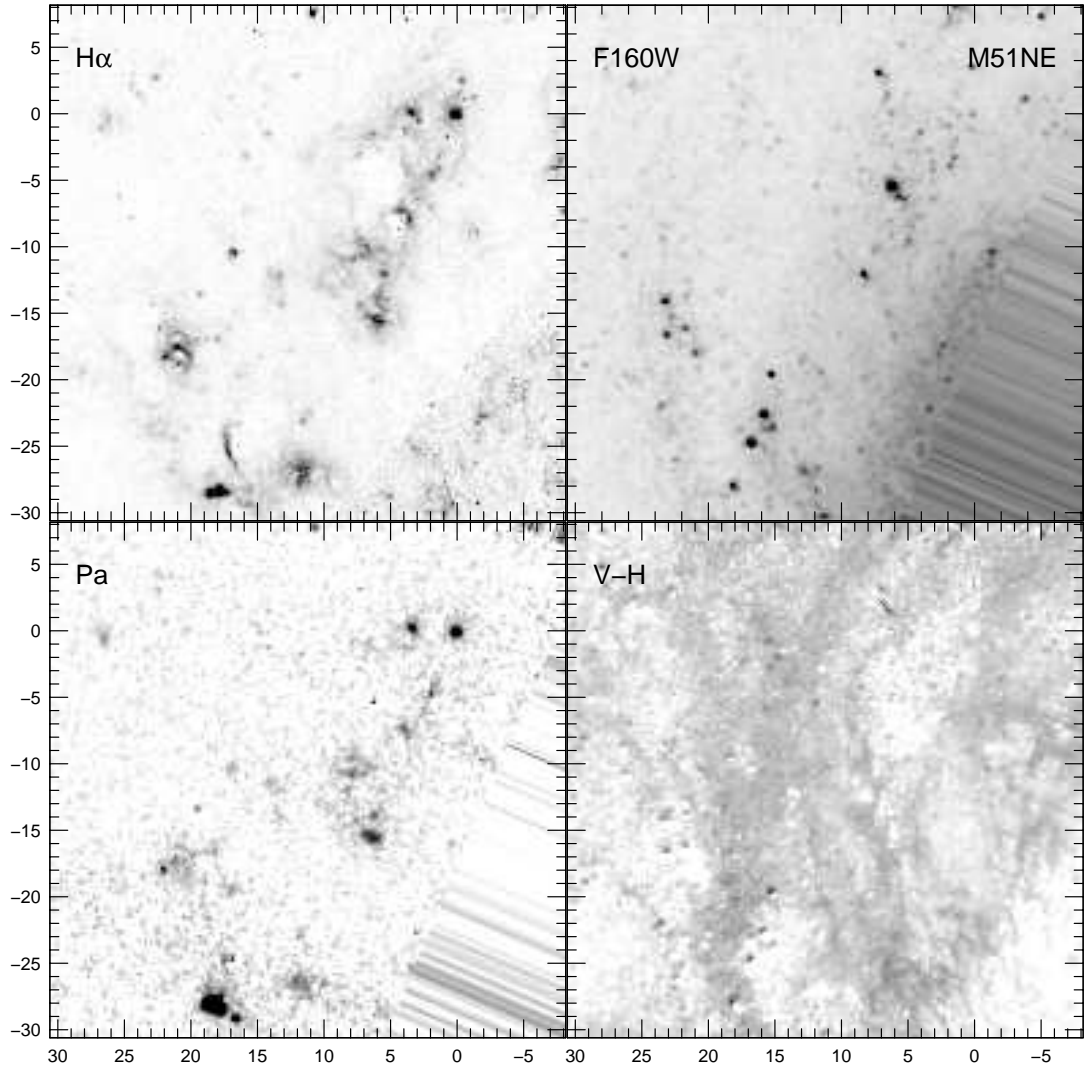}
\caption[junk]{
HST images of a region about $20''$ to the north and $20''$ to the 
east of the nucleus in M51. The H$\alpha$
image was constructed from WFPC2 images.  The Pa$\alpha$
was observed with NICMOS Camera 3 and constructed from F187N and F190N
images.  The V-H color map was constructed from the F547M and F160W
images.  Colors from V-H $= 2.5$ (white) to 3.3 (black) are shown.
Axes are given in arcsec from the brightest region listed in Table 6.
See Table 6 for the coordinates of this region.
}
\end{figure*}

\begin{figure*}
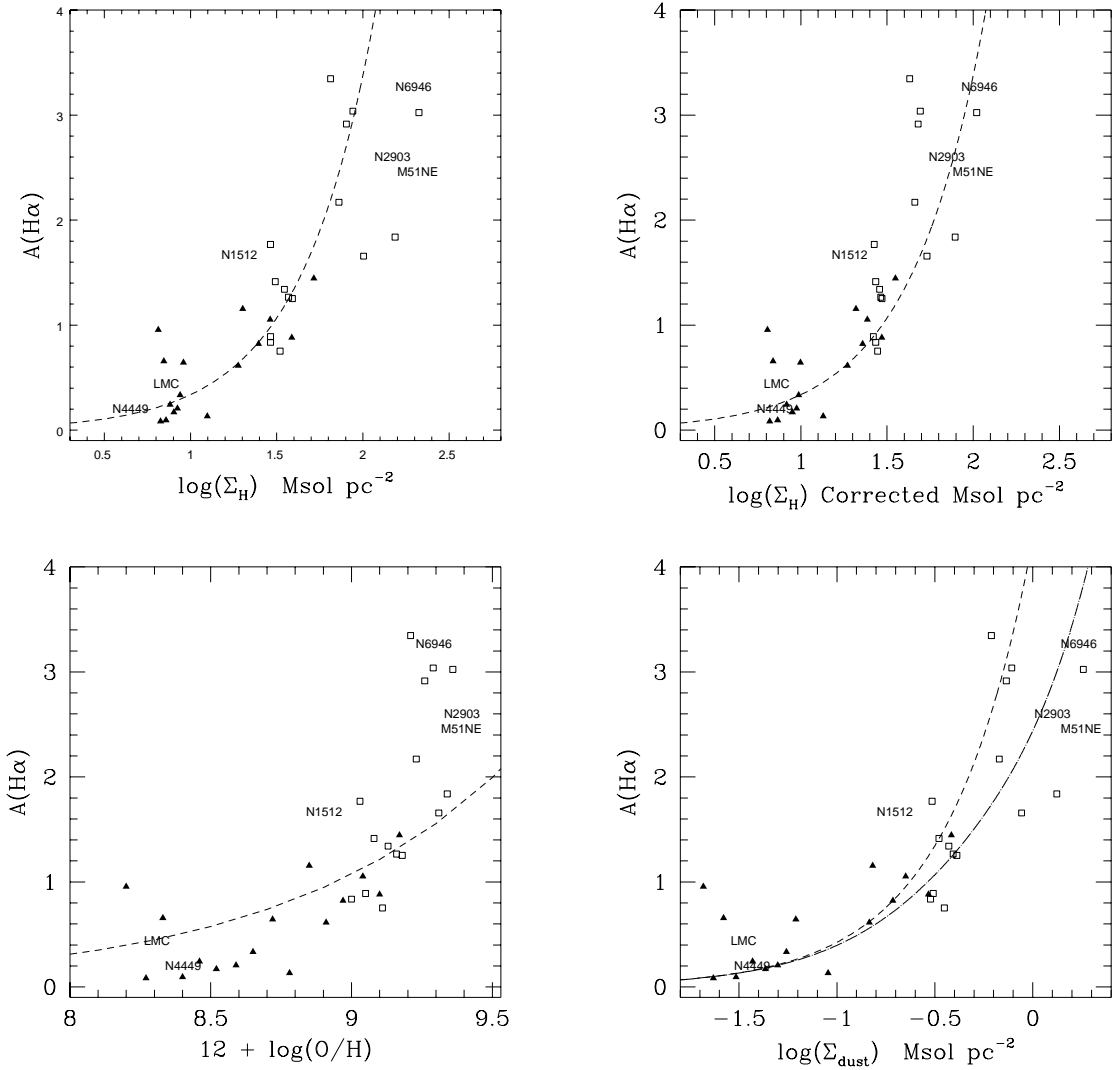

\vspace{13.0cm}
\includegraphics{fig4a.eps}
\includegraphics{fig4b.eps}
\includegraphics{fig4c.eps}
\includegraphics{fig4d.eps}
\caption[junk]{
\baselineskip=11pt
a) The relation between the  luminosity weighted extinction
and the azimuthal hydrogen gas density.
NGC 1512, NGC 2903, NGC~4449, NGC6946, the LMC, and the north
east position of M51 are plotted by name.
The M101 and M51 azimuthal averages are plotted as solid triangles and open
squares, respectively, where each point represents an average
at a different radius.
The dashed line shows the linear relation between $A_V$ and
half of the gas column depth (assuming the relation between
hydrogen column depth and extinction from Mathis 1990).
We use half of the gas column, assuming that the \ion{H}{2} regions are embedded
in the gas disk.
The gas densities have been estimated with a CO conversion factor
of N(H$_2$)~$= 2.8\times 10^{20} {\rm cm}^{-2} I_{CO}$ (K km s$^{-1}$)
(Scoville et al.~1987).
b)  Similar to a) however
we have corrected the hydrogen gas densities by using
a CO conversion factor that depends on metallicity (see text for details).
c) Dependence of the luminosity weighted extinction with metallicity only.
We see this correlation in part because the regions with
higher gas density also correspond to regions of higher metallicity.
The dotted line shows the extinction predicted from
a gas surface density of $10 M_\odot {\rm pc}^{-2}$
and a dust to gas fraction that depends on the carbon abundance.
The metallicity alone is not sufficient to account for the variations
in extinctions.
d) Dependence of the luminosity weighted extinction (defined in
Eqn 3) with estimated
dust surface density.  The dust surface density was estimated
from the corrected gas densities (shown in b) by scaling
with the carbon abundance (see text for details).
The dashed line shows the linear relation between $A_V$ and
half of the dust column depth.
The dot dashed line is a 2 component patchy dust model with an
area filling factor of 50\% and an opacity ratio of 4. This model
illustrates that a patchy medium would have a lower mean extinction
than that predicted directly from the dust density.
}
\end{figure*}

\end{document}


\pagestyle{empty}

\begin{deluxetable}{lccccccccccc}
\tablewidth{0pt}
\tablecaption{HST archival images}
\tablehead{                                     
\colhead{galaxy} &
\colhead{Pa$\alpha$} &
\colhead{continuum} &
\colhead{H$\alpha$} &
\colhead{continuum} 
}
\startdata
NGC~1512 &  F187N/NIC2 & F160W/NIC2 & F658N & F547M \\
NGC~2903 &  F187N/NIC3 & F160W/NIC3 & F656N & F555W \\
M51      &  F187N/NIC3 & F190N/NIC3 & F656N & F547M \\
NGC~4449 &  F187N/NIC3 & F160W/NIC3 &       &       \\
NGC~6946 &  F187N/NIC3 & F190N/NIC3 &       &       \\
\enddata
\tablecomments{
Line and continuum filters on HST used to make the Pa$\alpha$ and H$\alpha$
line emission maps shown in Figures 1-3 and used to measure \ion{H}{2} region
intensities and line ratios listed in Tables 2-6.  
The near-infrared images were observed
with NICMOS and the filter names are followed by the NICMOS camera used for
the observation.  
The pixel size for the NICMOS camera 2 (NIC2) is $\sim 0\farcs 076$,
and for the NICMOS camera 3 (NIC3) is $\sim 0\farcs 204$.
}
\end{deluxetable}

\begin{deluxetable}{lccccccccccc}
\tablewidth{0pt}
\tablecaption{\ion{H}{2} regions in NGC~1512}
\tablehead{
\colhead{$\Delta$ RA}  &
\colhead{$\Delta$ DEC} &
\colhead{Flux H$\alpha$}   &
\colhead{$F_{Pa\alpha}/F_{H\alpha}$}   &
\colhead{$A_V$}  &
\colhead{$L_{H\alpha}$}    \\
\colhead{$''$}            &
\colhead{$''$}            &
\colhead{$10^{-14}$ ergs cm$^{-2}$ s$^{-1}$}       &
\colhead{}                &
\colhead{}                &
\colhead{$10^{37}$ ergs s$^{-1}$ }            \\
\colhead{(1)}             & 
\colhead{(2)}             & 
\colhead{(3)}             & 
\colhead{(4)}             & 
\colhead{(5)}             & 
\colhead{(6)}             
}
\startdata
    1.29 &  -6.16  & 36.97 &   0.32  &  1.62  & 13.60 \\
    3.65 &  -5.79  & 10.83 &   0.40  &  1.98  &  5.22 \\
    3.88 &  -5.24  & 18.68 &   0.31  &  1.54  &  6.44 \\
    5.42 &  -5.09  & 10.21 &   0.41  &  2.01  &  5.02 \\
    5.42 &  -4.67  & 14.85 &   0.30  &  1.49  &  4.92 \\
    5.93 &  -4.71  &  6.67 &   0.52  &  2.42  &  4.46 \\
    7.51 &  -1.92  &  2.61 &   0.79  &  3.07  &  2.86 \\
    8.28 &  -1.54  &  5.41 &   0.36  &  1.82  &  2.31 \\
    7.30 &   2.58  &  4.39 &   0.33  &  1.66  &  1.66 \\
    6.76  &  3.22  &  5.35 &   0.39  &  1.92  &  2.46 \\
    5.40  &  4.51  & 12.24 &   0.17  &  0.57  &  2.03 \\
    4.72  &  4.41  &  1.75 &   0.55  &  2.51  &  1.25 \\
    3.81  &  4.94  &  3.41 &   0.43  &  2.09  &  1.78 \\
    2.48  &  5.40  &  2.53 &   0.90  &  3.28  &  3.27 \\
   -6.47  &  3.95  &  5.92 &   0.52  &  2.41  &  3.93 \\
   -6.76  &  3.57  &  8.11 &   0.63  &  2.70  &  6.75 \\
   -7.52  & -1.59  &  1.46 &   0.88  &  3.25  &  1.84 \\
   -6.61  & -3.26  &  6.48 &   0.75  &  2.99  &  6.70 \\
   -6.58  & -3.27  &  6.27 &   0.79  &  3.07  &  6.89 \\
\enddata
\tablecomments{
We did aperture photometry in $0\farcs 75$  radius apertures.  Offsets 
are given with respect to the position of the nucleus;
RA(2000) = 4:03:54.131, 
DEC(2000)= -43:20:56.54, as measured from the HST pointing.
Columns: (1) Offset in RA from the nucleus.  A positive offset 
is east of the nucleus; (2) Offsets in DEC from the nucleus.  
A positive offset is 
north of the nucleus; (3) Measured flux in H$\alpha$.
Uncertainty in the sky value causes an error of $\sim \pm 5\%$
in the faintest regions; 
(4) Ratio of flux in Pa$\alpha$ to H$\alpha$. 
Uncertainty in the sky value causes an error of $\sim \pm 13\%$
in the measurement of the Pa$\alpha$ flux in the faintest regions; 
(5) $A_V$ computed from the ratio.
The distribution of extinctions is centered at $A_V = 1.8$ with
a standard deviation of $0.7$;
(6) Luminosity in H$\alpha$ which has been corrected for extinction 
using $A_V$ given in column (5). 
We assume a distance of 9.5 Mpc 
(Tully 1988; $H_0=75$~km~s$^{-1}$~Mpc$^{-1}$).
}
\end{deluxetable}

\begin{deluxetable}{lccccccccccc}
\tablewidth{0pt}
\tablecaption{\ion{H}{2} regions in NGC~2903}
\tablehead{
\colhead{$\Delta$ RA}  &
\colhead{$\Delta$ DEC} &
\colhead{Flux H$\alpha$}   &
\colhead{$F_{Pa\alpha}/F_{H\alpha}$}   &
\colhead{$A_V$}  &
\colhead{$L_{H\alpha}$}    \\
\colhead{$''$}            &
\colhead{$''$}            &
\colhead{$10^{-14}$ ergs cm$^{-2}$ s$^{-1}$}       &
\colhead{}                &
\colhead{}                &
\colhead{$10^{38}$ ergs s$^{-1}$ }            \\
\colhead{(1)}             & 
\colhead{(2)}             & 
\colhead{(3)}             & 
\colhead{(4)}             & 
\colhead{(5)}             & 
\colhead{(6)}             
} 
\startdata
   -2.95  &   2.79  &   1.79  &   0.91  &   3.21  &   9.56 \\
   -1.38  &   3.19  &   2.00  &   0.47  &   2.13  &   4.73 \\
   -1.11  &   4.02  &   2.50  &   0.47  &   2.12  &   5.87 \\
   -0.03  &   5.45  &   1.63  &   0.89  &   3.17  &   8.46 \\
    0.71  &   4.17  &   1.44  &   1.66  &   4.18  &  15.99 \\
    2.17  &   5.61  &   0.70  &   1.47  &   3.98  &   6.70 \\
    2.58  &   5.04  &   0.80  &   1.35  &   3.84  &   6.85 \\
    3.81  &   2.20  &   0.48  &   1.65  &   4.17  &   5.35 \\
    2.17  &   2.99  &   1.49  &   0.75  &   2.89  &   6.25 \\
    1.97  &   1.42  &   1.89  &   0.56  &   2.41  &   5.51 \\
    0.24  &  -2.66  &   0.75  &   1.04  &   3.42  &   4.71 \\
   -0.49  &  -3.36  &   1.23  &   0.81  &   3.02  &   5.68 \\
   -0.98  &  -3.98  &   0.55  &   1.48  &   4.00  &   5.29 \\    
\enddata
\tablecomments{
We did aperture photometry in $0\farcs 469$ radius apertures.  Offsets are given 
with respect to the position of the nucleus; RA(2000) = 9:32:09.98, 
DEC(2000)= +21:30:0.8, as measured from the HST pointing.
Columns: (1) Offset in RA from the nucleus.  A positive offset is east of 
the nucleus; (2) Offset in DEC from the nucleus.  A positive offset is north 
of the nucleus; (3) Measured flux in H$\alpha$. 
Uncertainty in the sky value causes an error of $\sim \pm 2\%$
in the faintest regions; 
(4) Ratio of flux in Pa$\alpha$ to H$\alpha$. 
Uncertainty in the sky value causes an error of $\sim \pm 4\%$
in the measurement of the Pa$\alpha$ flux in the faintest regions; 
(5) $A_V$ computed from the ratio. 
The distribution of extinctions is centered at $A_V = 3.3$ with
a standard deviation of $0.7$;
(6) Luminosity in H$\alpha$ which has been corrected
for extinction using  $A_V$ given in column (5). 
We assume a distance of 6.3 Mpc 
(Tully 1988; $H_0=75$~km~s$^{-1}$~Mpc$^{-1}$).
}
\end{deluxetable}

\begin{deluxetable}{lccccccccccc}
\tablewidth{0pt}
\tablecaption{\ion{H}{2} regions in NGC~4449}
\tablehead{
\colhead{RA(2000)}  &
\colhead{DEC(2000)} &
\colhead{Flux Pa$\alpha$}   &
\colhead{$F_{Pa\alpha}/F_{H\alpha}$}   &
\colhead{$A_V$}  &
\colhead{$L_{H\alpha}$}    \\
\colhead{$''$}            &
\colhead{$''$}            &
\colhead{$10^{-14}$ ergs cm$^{-2}$ s$^{-1}$}       &
\colhead{}                &
\colhead{}                &
\colhead{$10^{37}$ ergs s$^{-1}$ }            \\
\colhead{(1)}             & 
\colhead{(2)}             & 
\colhead{(3)}             & 
\colhead{(4)}             & 
\colhead{(5)}             & 
\colhead{(6)}             
} 
\startdata
  12:28:12.08 & 44:05:51.56 &  1.49 &  0.05 &  0.00 & 3.39 \\
  12:28:11.70 & 44:05:55.98 &  0.82 &  0.02 &  0.00 & 3.67 \\
  12:28:11.33 & 44:05:42.12 &  1.40 &  0.02 &  0.00 & 7.64 \\
  12:28:11.83 & 44:05:46.18 &  2.35 &  0.07 &  0.00 & 3.88 \\
  12:28:12.93 & 44:05:31.61 &  2.02 &  0.11 &  0.00 & 1.98 \\
  12:28:13.68 & 44:05:42.98 &  1.18 &  0.07 &  0.00 & 1.98 \\
  12:28:13.11 & 44:05:41.44 &  5.28 &  0.29 &  1.43 & 6.04 \\
  12:28:12.45 & 44:05:45.13 &  1.50 &  0.05 &  0.00 & 3.09 \\
  12:28:13.74 & 44:06:10.35 &  2.18 &  0.12 &  0.07 & 2.06 \\
  12:28:12.63 & 44:05:21.34 &  1.75 &  0.10 &  0.00 & 2.01 \\
  12:28:13.70 & 44:05:26.55 &  0.82 &  0.13 &  0.16 & 0.79 \\
  12:28:11.37 & 44:05:36.89 & 12.79 &  0.19 &  0.73 & 13.27 \\    
\enddata
\tablecomments{
We did aperture photometry in $4''$ radius apertures to match the apertures 
used by Scowen (1992) to measure H$\alpha$ fluxes.
Columns: (1) RA of emission peak in RA as we measure in the Pa$\alpha$ image; 
(2) DEC of emission peak as measured in the Pa$\alpha$ image;
(3) Measured flux in Pa$\alpha$. 
Uncertainty in the sky value causes an error of $\sim \pm 11\%$
in the measurement of the Pa$\alpha$ flux in the faintest regions; 
(4) Ratio computed by using our 
Pa$\alpha$ flux and the H$\alpha$ fluxe measured by Scowen (1992) 
in the same region;
(5) $A_V$ computed from the ratio; (6) Luminosity in H$\alpha$
which has been corrected for 
extinction using given $A_V$ given in column (5).  
We assume a distance of 3.05 Mpc (Stewart et al.~2000).
}
\end{deluxetable}

\begin{deluxetable}{lccccccccccc}
\tablewidth{0pt}
\tablecaption{\ion{H}{2} regions in NGC~6946}
\tablehead{
\colhead{$\Delta$ RA}  &
\colhead{$\Delta$ DEC} &
\colhead{Flux Pa$\alpha$}   &
\colhead{$F_{Pa\alpha}/F_{H\alpha}$}   &
\colhead{$A_V$}  &
\colhead{$L_{H\alpha}$}    \\
\colhead{$''$}            &
\colhead{$''$}            &
\colhead{$10^{-14}$ ergs cm$^{-2}$ s$^{-1}$}       &
\colhead{}                &
\colhead{}                &
\colhead{$10^{38}$ ergs s$^{-1}$ }            \\
\colhead{(1)}             & 
\colhead{(2)}             & 
\colhead{(3)}             & 
\colhead{(4)}             & 
\colhead{(5)}             & 
\colhead{(6)}             
} 
\startdata
    0.75 &  -1.49 &  28.20 &   2.03 &   4.60 & 163.16 \\
   -1.24 & -15.97 &   1.28 &   0.47 &   2.24 &   5.34 \\
  -13.48 & -10.16 &   1.74 &   0.86 &   3.21 &   8.30 \\
   -9.00 & -23.62 &   3.50 &   0.82 &   3.15 &  16.57 \\
  -11.60 &   2.28 &   2.65 &   0.85 &   3.20 &  12.64 \\
   16.36 & -22.69 &   1.29 &   0.55 &   2.48 &   5.55 \\    
\enddata
\tablecomments{
We did aperture photometry in $2\farcs 7$ radius apertures.  
Offset are given with respect to the position of the nucleus; 
RA(2000) = 20:34:52.33, DEC(2000)= +60:09:14.21, 
as measured from the HST pointing.
Columns: (1) Offset in RA from the nucleus.  A positive offset is east 
of the nucleus; (2) Offset in DEC for the nucleus.  A positive offset is 
north of the nucleus; (3) Measured flux in Pa$\alpha$. 
Uncertainty in the sky value causes an error of $\sim \pm 11\%$
in the measurement of the Pa$\alpha$ flux in the faintest region; 
(4) Ratio of flux in Pa$\alpha$ to H$\alpha$. 
Uncertainty in the sky value causes an error of $\sim \pm 15\%$
in the measurement of the H$\alpha$ flux in the faintest region; 
(5) $A_V$ computed from the ratio. 
The distribution of extinctions is centered at $A_V = 2.8$ with
a standard deviation of $1.2$;
(6) Luminosity in H$\alpha$ which has been corrected for extinction 
using $A_V$ given in column (5).
We assume a distance of 5.5 Mpc 
(Tully 1988; $H_0=75$~km~s$^{-1}$~Mpc$^{-1}$).
}
\end{deluxetable}

\begin{deluxetable}{lccccccccccc}
\tablewidth{0pt}
\tablecaption{\ion{H}{2} regions in M51}
\tablehead{
\colhead {$\Delta$ RA}  &
\colhead{$\Delta$ DEC} &
\colhead{Flux H$\alpha$}   &
\colhead{$F_{Pa\alpha}/F_{H\alpha}$}   &
\colhead{$A_V$}  &
\colhead{$L_{H\alpha}$}    \\
\colhead{$''$}            &
\colhead{$''$}            &
\colhead{$10^{-15}$ ergs cm$^{-2}$ s$^{-1}$}       &
\colhead{}                &
\colhead{}                &
\colhead{$10^{38}$ ergs s$^{-1}$ }            \\
\colhead{(1)}             & 
\colhead{(2)}             & 
\colhead{(3)}             & 
\colhead{(4)}             & 
\colhead{(5)}             & 
\colhead{(6)}             
} 
\startdata
  0.00 &   0.00 & 12.85 &  0.30 &  1.51 &  5.23 \\
  3.29 &   0.39 &  4.99 &  0.51 &  2.38 &  3.88 \\
  1.84 &  -4.58 &  3.01 &  0.42 &  2.05 &  1.83 \\
  3.86 &  -7.24 &  3.29 &  0.41 &  2.02 &  1.95 \\
  7.81 & -10.60 &  2.11 &  0.65 &  2.77 &  2.20 \\
  6.12 & -13.77 &  2.45 &  0.54 &  2.45 &  2.02 \\
  6.24 & -15.54 &  3.99 &  0.58 &  2.59 &  3.62 \\
 12.04 & -26.23 &  3.27 &  0.45 &  2.18 &  2.19 \\
 18.15 & -27.76 &  3.89 &  2.61 &  5.01 & 21.97 \\
 16.51 & -29.02 &  1.37 &  2.40 &  4.88 &  6.95 \\
\enddata
\tablecomments{
We did aperture photometry in $0\farcs 41$ radius apertures.  Offsets are given 
with respect to the first listed position; 
RA(2000) = 13:29:54.00, DEC(2000) = 47:12:11.86.
Columns : (1) Offset in RA from the first listed position.  
A positive offset is east 
of the first position; (2) Offset in DEC from the first 
position.  A positive offset is 
north of the first detected position nucleus;  
(3) Measured flux in H$\alpha$. 
Uncertainty in the sky value causes an error of $\sim \pm 4\%$
in the measurement of the H$\alpha$ flux in the faintest region; 
(4) Ratio of flux in Pa$\alpha$ to H$\alpha$. 
Uncertainty in the sky value causes an error of $\sim \pm 8\%$
in the measurement of the H$\alpha$ flux in the faintest region; 
(5) $A_V$ computed from the ratio; 
(6) Luminosity in H$\alpha$ which has been corrected for extinction using  
$A_V$ given in column (5).
The assumed distance is 9.6 Mpc (Sandage \& Tammann 1975).
}
\end{deluxetable}

\begin{deluxetable}{lccccccccccc}
\tablewidth{0pt}
\tablecaption{Integrated properties of \ion{H}{2} regions, local gas density
and metalicity}
\tablehead{
\multicolumn{1}{l}{Galaxy}  &
\colhead{$\overline{A(H \alpha)}$}  &
\colhead{$\overline{A(Pa\alpha)}$}  &
\colhead{$A_{total}(H\alpha)$}      &
\colhead{12 + log(O/H)}  &
\colhead{$M_{HI}$}         &
\colhead{$M_{H2}$}         \\
\colhead{}         &
\colhead{}         &
\colhead{}         &
\colhead{}         &
\colhead{}         &
\colhead{$M_\odot$ pc$^{-2}$}         &
\colhead{$M_\odot$ pc$^{-2}$}         \\
\colhead{(1)}       &
\colhead{(2)}       &
\colhead{(3)}       &
\colhead{(4)}       &
\colhead{(5)}       &
\colhead{(6)}       &
\colhead{(7)}       
}
\startdata
NGC~1512 & 1.68 & 0.32    & 1.67   & 9.4  &      & 19   \\
NGC~2903 & 2.62 & 0.48    & 2.67   & 8.9  &      & 147  \\
NGC~4449 & 0.25 & 0.07    & 0.00   & 8.4  & 2.2  & 2.2  \\
NGC~6946 & 3.28 & 0.61    & 3.26   & 9.3  &      & 195  \\
M51NE    & 2.48 & 0.52    & 2.37   & 9.4  &      & 200  \\
LMC      & 0.45 &         &        & 8.8  & 7    &      \\
\enddata
\tablecomments{
Values for the mean extinctions $\overline{A(H\alpha)}$ and 
$\overline{A(Pa\alpha)}$  
(defined in Equation 3) were derived from the Pa$\alpha$
fluxes and Pa$\alpha$/H$\alpha$ flux ratios listed in Tables 2-6
for all galaxies except the LMC.
This can be compared to the extinction estimated from the ratio of
the total Pa$\alpha$ and H$\alpha$ fluxes which we
refer to as $A_{total}(H\alpha)$.
The value for $\overline{A(H\alpha)}$ for the LMC was estimated from
the fluxes and Balmer decrements listed by Caplan \& Deharveng (1985).
Metalicities are estimated for the region covered by our images.   
For NGC~1512 the central metalicity was estimated
based on its luminosity and a gas density corresponding to the extinction
we observe in our color map (see text).  Central metalicities
for NGC~2903, M51, and NGC~6494 were estimated from gradients
and values listed by Zaritsky et al.~(1994).
The metalicity of NGC~4449 is that measured by Heckman et al.~(1988).
The gas density for the nucleus of NGC~2903 was estimated from
the inclination corrected value by Jackson et al.~(1991).
The gas density for NGC~6946 is that listed by Young et al~(1995)
and is averaged over the NICMOS Camera 3 field of view ($\sim 45''$).
The gas densities for NGC~4449 
are estimated using an average of values listed in 
Hunter \& Thronson (1996).
Gas densities for the LMC were taken from Luks \& Rohlfs (1992)
and Cohen et al.~(1988).
The molecular gas density for the NE position of M51 was taken
from Kuno \& Nakai (1997).
When necessary we have corrected published values so that we used
the same conversion factor of
N(H$_2$)~$= 2.8\times 10^{20} {\rm cm}^{-2} I_{CO}$~(K~km~s$^{-1}$)
(Scoville et al.~1987).
}
\end{deluxetable}